\documentclass{article}

\usepackage{PRIMEarxiv}

\usepackage{array}

\usepackage[utf8]{inputenc} 
\usepackage[T1]{fontenc}    
\usepackage{hyperref}       
\usepackage{url}            
\usepackage{booktabs}       
\usepackage{amsfonts}       
\usepackage{nicefrac}       
\usepackage{microtype}      
\usepackage{lipsum}
\usepackage{fancyhdr}       
\usepackage{graphicx}       
\graphicspath{{media/}}     
\usepackage{algorithm} 
\usepackage{algorithmic} 
\usepackage{color}
\usepackage{arydshln} 
\usepackage{amsfonts} 
\usepackage{amsmath}

\title{Enhancing Inverse Problem Solutions with Accurate Surrogate Simulators and Promising Candidates
}

\author{
  Akihiro Fujii\\
  Department of Materials Engineering\\
  University of Tokyo \\
  \texttt{akihiro.fujii@cello.t.u-tokyo.ac.jp} \\
   \And
  Hideki Tsunashima \\
  School of Advanced Science and Engineering\\
  Waseda University  \\
  \texttt{h.tsunashima@asagi.waseda.jp} \\
   \And
  Yoshihiro Fukuhara \\
  ExaWizards Inc.  \\
  \texttt{yoshihiro.fukuhara@exwzd.com} \\
   \And
  Koji Shimizu \\
  Department of Materials Engineering\\
  University of Tokyo \\
  \texttt{shimizu@cello.t.u-tokyo.ac.jp} \\
   \And
  Satoshi Watanabe \\
  Department of Materials Engineering\\
  University of Tokyo \\
  \texttt{watanabe@cello.t.u-tokyo.ac.jp} \\
}

\begin{document}
\maketitle

\begin{abstract}
Deep-learning inverse techniques have attracted significant attention in recent years. Among them, the neural adjoint (NA) method, which employs a neural network surrogate simulator, has demonstrated impressive performance in the design tasks of artificial electromagnetic materials (AEM). However, the impact of the surrogate simulators’ accuracy on the solutions in the NA method remains uncertain. Furthermore, achieving sufficient optimization becomes challenging in this method when the surrogate simulator is large, and computational resources are limited. Additionally, the behavior under constraints has not been studied, despite its importance from the engineering perspective. In this study, we investigated the impact of surrogate simulators’ accuracy on the solutions and discovered that the more accurate the surrogate simulator is, the better the solutions become. We then developed an extension of the NA method, named Neural Lagrangian (NeuLag) method, capable of efficiently optimizing a sufficient number of solution candidates. We then demonstrated that the NeuLag method can find optimal solutions even when handling sufficient candidates is difficult due to the use of a large and accurate surrogate simulator. The resimulation errors of the NeuLag method were approximately 1/50 compared to previous methods for three AEM tasks. Finally, we performed optimization under constraint using NA and NeuLag, and confirmed their potential in optimization with soft or hard constraints. We believe our method holds potential in areas that require large and accurate surrogate simulators.
\end{abstract}


\section{Introduction}
In recent years, significant research has focused on approximating resource-intensive physics simulations using deep-learning techniques.
Deep learning has been employed to approximate various physics simulations, such as time-dependent dynamics in fluids \cite{raissi2019physics, mao2020physics}, the relationship between energy and atomic configuration (NNPs: Neural Network Potentials) \cite{hermann2020deep, pfau2020ab, behler2015constructing, schutt2017schnet}, and the connection between geometry and electromagnetic (EM) response in artificial electromagnetic materials (AEM) \cite{kuhn2022inverse, deng2021benchmarking}. Machine learning models, including neural networks that approximate physics simulators, are occasionally referred to as surrogate simulators.

Approximating a physics simulator is undoubtedly important; however, solutions to the opposite-direction problem, known as the inverse problem, are equally significant. AEM design, which aims to obtain a geometry that yields the desired EM response, is a prime example of such an inverse problem. \cite{staude2017metamaterial, khatib2021deep, liu2018generative, ma2018deep, peurifoy2018nanophotonic, acharige2022machine, estrada2022inverse, chen2019smart}. 
Solving inverse problems such as AEM design is often more challenging than their counterparts (i.e., “forward problems”) because, in the inverse problem, one input often corresponds to multiple outputs. 

Due to its practical importance, AEM design has been actively explored. Various deep inverse methods (DIMs), which are approaches for inverse problems using deep-learning techniques, have already been applied to this subject. These methods include conventional neural network-based approaches \cite{chen2019smart}, generative models-based methods (Autoencoder \cite{hinton2006reducing}, cVAE \cite{sohn2015learning}, and generative adversarial networks (GANs) \cite{goodfellow2020generative}) \cite{liu2018generative, so2019designing, qiu2019deep}, Tandem Network (TD) based method \cite{liu2018training}, Invertible Neural Network (INN) \cite{ardizzone2018analyzing} and conditional INN (cINN) \cite{kruse2021benchmarking} based methods, and Mixture Density Network (MDN) \cite{bishop1994mixture} based methods \cite{unni2021mixture, unni2020deep}. Additionally, some methods combine neural networks and genetic algorithms (GA) \cite{da2014optimization, liu2020hybrid}. 

Among these methods, The Neural Adjoint (NA) method \cite{ren2020benchmarking} has garnered significant attention due to its outstanding results in inverse-problem benchmarking. 
The only architectural requirement for the surrogate simulator in the NA method is the applicability of the backpropagation technique. As a result, it does not necessitate an invertible function, which is essential in INNs, allowing state-of-the-art methods to be employed for approximating physical simulators using neural networks. Additionally, the NA method can address one-to-many problems by simultaneously optimizing many solution candidates. Because of these properties, the NA method is promising for inverse physics problems and has achieved excellent results in AEM design tasks \cite{ren2022inverse}. 

However, it remains uncertain whether adopting a more accurate surrogate model can yield better solutions to inverse problems. Furthermore, behavior under constraints has not been studied, despite its importance from an engineering perspective.

In this study, we first examined the relationship between the accuracy of the surrogate simulator and the quality of the solution. We confirmed that a more accurate surrogate simulator leads to better solutions. We also demonstrated that using a highly accurate surrogate simulator significantly increases computational costs, thereby limiting the usefulness of the NA method. To address the trade-off between quality of solutions and computational cost, we proposed a new method called the Neural Lagrangian (NeuLag) method. Similar to the NA method, it optimizes solution candidates but employs a larger number of candidates and focuses on the promising ones.

We subsequently applied the NueLag method to three AEM tasks (Stack, Shell, and ADM, which will be explained later) and compared its performance with those of previous methods, including the NA method. The NeuLag method not only succeeded in reducing the resimulation error (evaluation metric for the inverse problem, described in 4.1.1) of the previous method from 1/5 to 1/50 but also achieved approximately three times faster convergence speed for optimization than the NA method in a scenario with a limited batch size due to a highly accurate but large surrogate simulator.

Additionally, we demonstrateddds how to introduce hard or soft constraints into the NA and NeuLag methods. Hard constraints force a feature to a specific value, whereas soft constraints use a loss function to provide lenient constraints on the solution candidates. We evaluated the NA and NeuLag methods under two types of constraints and investigated their behavior.

The code is available at \url{https://github.com/AkiraTOSEI/neulag}.

\section{In-depth review of neural adjoint method}
\label{rivisit_NA}

In this section, we first outline the NA method algorithm in Section \ref{NA_method} and then investigate its properties. In Section \ref{UB_accumary}, we demonstrate that the accuracy of the surrogate simulator firmly constrains the upper bound of solution quality. In Section \ref{num_candidates}, we reveal that employing a large number of solution candidates, widely distributed across the input space, is essential due to the one-to-many nature of AEM tasks.

\subsection{Algorithm of the NA method}
\label{NA_method}

The NA method is a two-step process. First, a surrogate simulator $\hat{f}_s$ is trained using standard supervised learning techniques to approximate the ground-truth simulator $f$ and solve the forward problem ($x$: input $\rightarrow y$: output). Using the mean squared error (MSE) as a loss function and splitting the dataset into training and validation sets are common practices in the field. Given that Ren et al.\cite{ren2022inverse} adopted the same approach, we decided to follow this method in our study. In the next step, the set of trainable solution candidates $\{\hat{x}_{c_1},... ,\hat{x}_{c_i},... ,\hat{x}_{c_n} \}$ is initialized and optimized iteratively using backpropagation to produce the desired output value $y_{gt}$. Notably, $y_{gt}$ is not a part of the training or validation data. The detailed procedure is presented in Algorithm \ref{alg1}. In the following section, we explain the algorithm in detail.

\begin{algorithm}[t]
\caption{NA method}
\label{alg1}
\begin{algorithmic}
\renewcommand{\algorithmicrequire}{\textbf{Input:}}
\renewcommand{\algorithmicensure}{\textbf{Output:}}
\REQUIRE target spectrum: $y_{gt}$, surrogate simulator:$\hat{f}_s$, learning rate: $\eta $
\ENSURE optimized solutions $\hat{X}_c$
\\ \textit{Initialization} : $\hat{X}_c = \{\hat{x}_{c_1},... ,\hat{x}_{c_n} \}$
\WHILE{$not \ converged$}
    \FOR{each $ \hat{x}_{c_i} \in \hat{X}_c $}
        \STATE $\hat{x}_{c_i} \leftarrow \hat{x}_{c_i} - \eta  \frac{\partial \mathrm{L}_{\mathrm{NA}}(\hat{x}_{c_i}, y_{gt})}{\partial \hat{x}_{c_i}} $
    \ENDFOR
\ENDWHILE
\end{algorithmic}
\end{algorithm}

First, we initialize $n$ of solution candidates $\hat{X}_c = \{\hat{x}_{c_1},... ,\hat{x}_{c_n} \}$ and obtain the output $\hat{y}$ using the surrogate simulator $\hat{f}_s$ for each $\hat{x}_{c_i}$. In AEM design tasks, the output is a spectrum. Then, we measure the distance between spectrum $\hat{y}$ and the target spectra $y_{gt}$. Various functions can be used as the distance function (or "loss function"), $\mathrm{L}_{\mathrm{FW loss}}(\hat{x}_{c_i},y_{gt})$. In this case, we employ the MSE, as used in the original study on the NA method\cite{ren2020benchmarking}.

\begin{equation}
\hat{f}_s(\hat{x}_{c_i}) = \hat{y}
\end{equation}

\begin{equation}
\label{fw_loss}
\mathrm{L}_{\mathrm{FW loss}}(\hat{x}_{c_i},y_{gt}) = \mathrm{MSE}(\hat{f}_s(\hat{x}_{c_i}), y_{gt})
\end{equation}

When optimizing using this loss, the solution candidate $\hat{x}_{c_i}$ can fall outside the data distribution. To mitigate this situation, the NA method penalizes the solutions that fall outside the data distribution. The loss added for this purpose, called the boundary loss $\mathrm{L}_{\mathrm{bnd}}$, can be written as follows using the mean $\mu_x$ and value range $R_x$ of the training data:

\begin{equation}
\mathrm{L_{\mathrm{bnd}}}(\hat{x}_{c_i}) = \mathrm{ReLU} ( |\hat{x}_{c_i} - \mu_x| - \frac{1}{2}R_x)
\end{equation}

This result can be interpreted as the introduction of soft constraints. The overall loss in the NA method is expressed as follows:

\begin{equation}
\mathrm{L_{\mathrm{NA}}}(\hat{x}_{c_i},y_{gt}) = \mathrm{L}_{\mathrm{FW loss}}(\hat{x}_{c_i},y_{gt}) + \mathrm{L_{\mathrm{bnd}}}(\hat{x}_{c_i})
\end{equation}

We use $\mathrm{L}_{\mathrm{NA}}$ to update each solution candidate $\hat{x}_{c_i}$ in each step. The update from step $t$ to step $t+1$ is expressed as follows, where $\eta$ is the learning rate:

\begin{equation}
\hat{x}^{t+1}_{c_i} = \hat{x}^t_{c_i} - \eta \left. \frac{\partial \mathrm{L}_{\mathrm{NA}}}{\partial x} \right|_{x=\hat{x}^t_{c_i}}
\end{equation}

In practice, optimization is performed in parallel for each batch, which is a set of solution candidates. This procedure is suitable for one-to-many problems that involve numerous local optima. By increasing the batch size, we are more likely to obtain initial values that converge to the global optimum.

\subsection{Upper bound of the solutions' quality}
\label{UB_accumary}

In general, the surrogate simulator $\hat{f}_s$ is not a perfect approximation of the ground-truth simulator $f$ for solving the forward problem in the NA method. One question arises: \textit{How does the approximation accuracy of the surrogate simulator $\hat{f}_s$ affect the solution quality of the inverse problem?} To answer this question, we prepared surrogate simulators with three levels of accuracy and investigated the relationship between the error predicted by the surrogate simulator $\hat{f}_s$ of the solution ($\mathrm{L}_{\mathrm{FW loss}}$) and the resimulation error of the ground-truth simulator $f$ ($\mathrm{L}_{\mathrm{resim}})$. The resimulation error $\mathrm{L}_{\mathrm{resim}}$ is expressed as follows:

\begin{equation}
\label{resim_error}
\mathrm{L}_{\mathrm{resim}}(\hat{x}_{c_i},y_{gt}) = \mathrm{MSE}(f(\hat{x}_{c_i}), y_{gt})
\end{equation}

\begin{figure}[t]
\centering
  \includegraphics[width=0.75\textwidth]{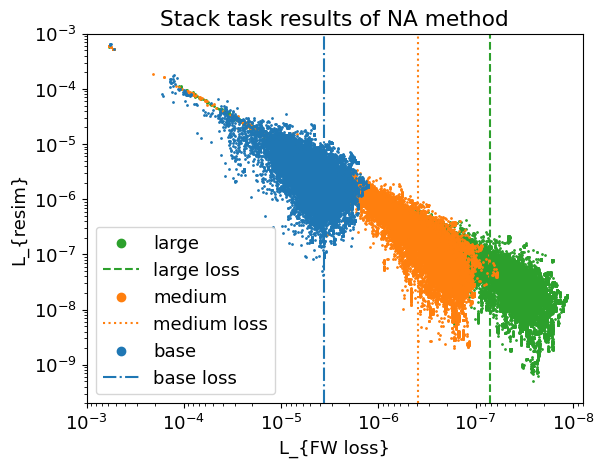}
  \caption{Relationship between surrogate simulator error ($\mathrm{L}_{\mathrm{FW loss}}$) and resimulation error ($\mathrm{L}_{\mathrm{resim}}$) for results of the Stack task with NA method. Three surrogate simulators with different accuracy, \textit{base}, \textit{medium}, and \textit{large}, were used in the experiments. Each point represents the solutions, and dashed lines represent the validation loss of surrogate simulators.}
  \label{fgr:acc_vs_inv}
\end{figure}

As shown in Fig.~\ref{fgr:acc_vs_inv}, we solved the inverse problem for the Stack task\cite{chen2019smart} using three surrogate simulators with different accuracies: \textit{base}, \textit{medium}, and \textit{large}. These models are all composed of residual blocks\cite{he2016deep}. While they share the same architectural components, they differ in terms of network depth and width. It is noteworthy that we employ a different structure from Ren \textit{et al.}\cite{ren2022inverse} to appropriately scale the surrogate simulator. For a comprehensive understanding of their configurations, please refer to Section \ref{surrogate_simulator_explain}.

The Stack task focuses on optimizing the geometry $x$ of a multilayer stack consisting of alternating graphene and Si3N4 dielectric layers to yield a target absorption spectrum $y$. Here, $x$ represents a 5-dimensional input, while $y$ corresponds to a 256-dimensional output. Comprehensive details about this task are elucidated in Table~\ref{tbl:dataset} and Section \ref{dataset}. 

As shown in Fig.~\ref{fgr:acc_vs_inv}, the proportional relationship between the losses of the surrogate simulators ($\mathrm{L}_{\mathrm{FW loss}}$) and the resimulation error ($\mathrm{L}_{\mathrm{resim}}$) breaks down around the validation loss of each surrogate simulator, indicating an upper bound on the solution quality (lower bound on the resimulation error $\mathrm{L}_{\mathrm{resim}}$) of the inverse problem.
In other words, inverse-problem methods using surrogate simulators, such as the NA method, require a highly accurate surrogate simulator to improve the solution quality. This is discussed in more detail in Section 4.2.

\subsection{Importance of having a large number of solution candidates}
\label{num_candidates}

The three AEM tasks of Stack, Shell, and ADM, which are described in Section~\ref{dataset}), are one-to-many problems \cite{ren2022inverse}. Because there are many local optima in a one-to-many problem, the solution candidates often fall into local optima and fail to reach the global optimum.

\begin{figure}[h]
\centering
  \includegraphics[width=\textwidth]{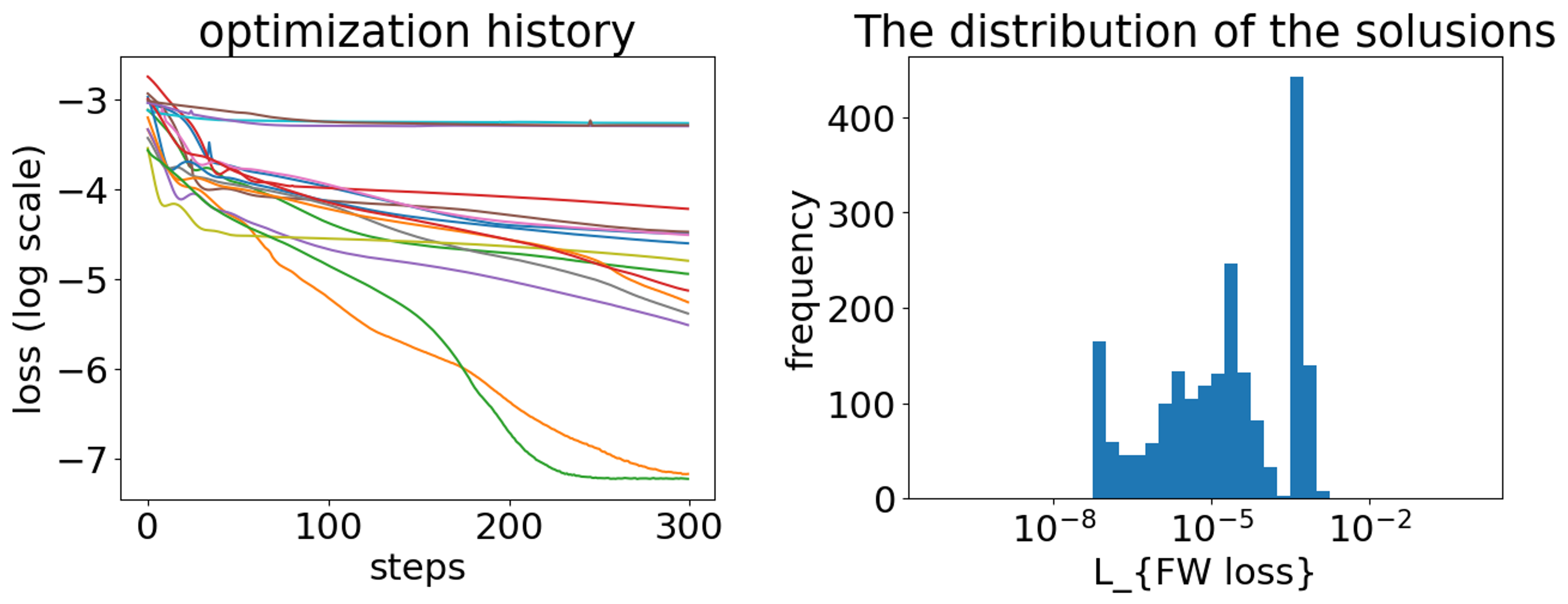}
  \caption{Qualities of optimized solution candidates. (left) Optimization histories of each solution candidates in the NA method. Each line represents one solution candidate. Some candidates fall into local optima, and their $\mathrm{L}_{\mathrm{FW loss}}$ does not decrease any further. (right) Final resimulation error distribution of optimized NA solutions. The distribution range is quite wide.}
  \label{fgr:candidate_distribution}
\end{figure}

In practice, to suppress such failures, the NA method initializes and optimizes as many solution candidates as the batch size. An example of behaviors of the solution candidates in the optimization of the NA method is shown in Fig.~\ref{fgr:candidate_distribution}. This figure clearly indicates that not all solution candidates reach the global optimum. Therefore, a large number of widely distributed solution candidates is highly desirable for reaching the global optimum.

\subsection{Difficulty in the NA method}
\label{difficulty_NA}

The discussion in Section~\ref{UB_accumary} suggests that adopting a highly accurate surrogate simulator is desirable. In the field of deep learning, there is a tendency for more accurate models to be larger. Such large models inherently consume a large amount of memory. Consequently, this also restricts the batch size, which, in the NA method, is directly related to the number of solution candidates.
Meanwhile, the discussion in Section~\ref{num_candidates} indicates that having several solution candidates is strongly desirable for reaching the global optimum.
Therefore, using large surrogate simulators in the NA model may degrade the solution quality.
The difficulty of simultaneously satisfying these two conditions limits the performance of the NA method.

\section{Neural Lagrangian method}
\label{neural_lagrangian}

To overcome the difficulty described in Section~\ref{difficulty_NA}, we propose a new method, named Neural Lagrangian (NeuLag) method. The name is derived from the form of loss function when introducing a constraint condition, which will be explained in Section~\ref{intro_const}. The NeuLag method is an extension of the NA method. 
In NeuLag method, we first use a highly accurate surrogate simulator to push the upper bound of the optimized solutions' quality.
Second, we focus on optimizing the promising solution candidates from among the large number of solution candidates. 
Owing to this strategy, we can obtain optimal solutions efficiently, even when batch size is small due to the use of a large surrogate simulator.

\subsection{Focusing on promising candidates}

As discussed in Section \ref{rivisit_NA}, only a few solution candidates using the NA method can reach the global optimum. Therefore, it is highly desirable to reduce the  the amount of the computational resources and memory used by the candidates that cannot reach the global optimum. To address this issue, we first halt the optimization, except for promising solution candidates, in the middle of the optimization. We then encourage further search for the remaining promising solution candidates. A Python-like pseudo algorithm is presented in Algorithm \ref{alg2}.

First, we initialize and optimize multiple solution candidates, as in the NA method. Next, based on the accumulated losses of each solution candidate, we retain $N$ promising candidates with small losses. There are various methods for accumulating losses such as simple summing.
We take the logarithm of the losses and use an exponential moving average (EMA) to reduce the effect of large losses in the initial steps. Finally, we \textit{branch out} the promising solution candidates. This process is described in detail in the following section.

\begin{algorithm}[t]
\caption{NeuLag method}
\label{alg2}
\begin{algorithmic}[1]
\renewcommand{\algorithmicrequire}{\textbf{Input:}}
\renewcommand{\algorithmicensure}{\textbf{Output:}}
\REQUIRE target spectrum: $y_{gt}$,  surrogate simulator:$\hat{f}_s$, learning rate: $\eta $, an activation function $\sigma$ \\ the number of final candidates: $N$, branches: $K$, NA optimization steps: NA\_Steps.
\ENSURE optimized solution $ \hat{X}_{c}$, optimized projector $g_{\theta}$
\\ \textit{Initialization} :
\\ \qquad $\hat{X}_{c_j} = \{\hat{x}_{c_1},... ,\hat{x}_{c_n} \}, g_{\theta}$
\\ \qquad $history = \mathrm{zero \_ tensor(shape=}(N,))$
\\ \qquad $Z_{c_i} = \{ z^j_{c_i} | j=1,...,K\}$,  $Z = \{ Z_{c_i} | i=1,...,N\}$
\\ \qquad $\Gamma_{c_i} = \{ \gamma^j_{c_i} | j=1,...,K\}$,  $\Gamma= \{ \Gamma_{c_i} | i=1,...,N\}$
\STATE \COMMENT{!<same procedure as NA method>}
\FOR{ NA\_Steps}
    \FOR{each $ \hat{x}_{c_i} \in \hat{X}_{c_j}$}
        \STATE $\hat{x}_{c_i} \leftarrow \hat{x}_{c_i} - \eta  \frac{\partial \mathrm{L}_{\mathrm{NL}}(y_{gt} \ , \ x_{c_i})}{\partial x} $
        \STATE $history[i] \leftarrow \mathrm{EMA} (history[i] + \mathrm{L}_{\mathrm{NL}}(x_{c_i} , y_{gt})) $
    \ENDFOR
\ENDFOR

\STATE
\STATE \COMMENT{!<candidate selection>}
\STATE best\_idx = argsort(history)[:N]
\STATE $\hat{X}_{c} \leftarrow \textbf{choose}$ best\_idx in $\hat{X}_{c}$

\STATE
\STATE \COMMENT{!<branching out candidates>}
\WHILE{$not \ converged$}
    \STATE L = []
    \FOR{each $ \hat{x}_{c_i} \in \hat{X}_{c}$}
        \FOR{each $z^j_{c_i} \in Z_{c_i}, \gamma^j_{c_i} \in \Gamma_{c_i}$}
            \STATE  $w \sim \mathcal{N}(0\ ,1)$
            \STATE $\hat{y} = \hat{f}_s (\sigma(x_{c_i} +\gamma^j_{c_i} g_{\theta}(z^j_{c_i}, w) ))$
            \STATE $loss = \mathrm{L}_{\mathrm{NL}}(\hat{y}, y_{gt})$
            \STATE $\hat{x}_{c_i} \leftarrow \hat{x}_{c_i} - \eta  \frac{\partial loss}{\partial \hat{x}_{c_i}} $
            \STATE $\gamma^j_{c_i} \leftarrow \gamma^j_{c_i} - \eta  \frac{\partial loss}{\partial \gamma^j_{c_i}} $
            \STATE $z^j_{c_i} \leftarrow z^j_{c_i} - \eta  \frac{\partial loss}{\partial z^j_{c_i}} $  
            \STATE L.append(loss)
        \ENDFOR
    \ENDFOR
    \STATE $\theta \leftarrow \theta - \eta  \frac{\partial \ mean(L)}{\partial \theta} $
\ENDWHILE

\end{algorithmic}
\end{algorithm}

\subsection{Branching out candidates with latent variables and an additional network}

\begin{figure}[t]
\centering
  \includegraphics[width=0.75\textwidth]{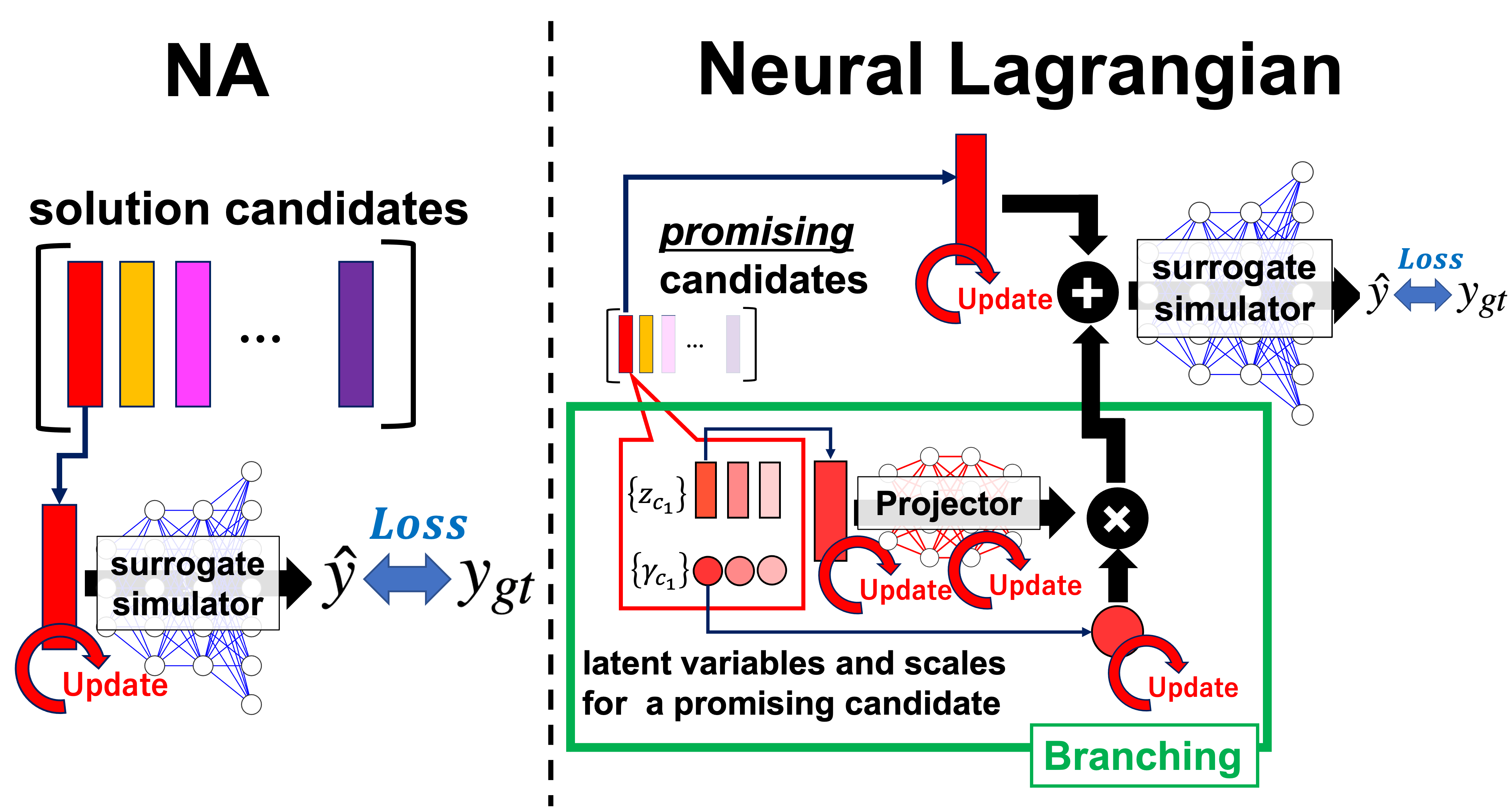}
  \caption{Conceptual diagrams of NA method and NeuLag method. (left) NA method. (right) Branching out candidates in NeuLag method. Promising solution candidates are branched out to search the optimal area more deeply using several $\gamma$ and $z$ per solution candidate. This procedure is described in lines 13 to 27 of Algorithm \ref{alg2}. }
  \label{fgr:concept}
\end{figure}

Here, we explain how to separate promising solution candidates (Fig \ref{fgr:concept}). We introduce several latent variables and a lightweight projector network $g_{\theta}$ that projects the latent variables to the input space. Specifically, we prepare K learnable latent variables $\{ z^{1}_{c_i},... ,z^{k}_{c_i}, ..., z^{K}_{c_i} \}$ for each promising solution candidate, $\hat{x}_{c_i}$.
We input these latent variables into the network $g_{\theta}$ and multiply the outputs by the coefficients $\gamma^{j}_{c_i}$. We then use the resulting values to modify the value of $\hat{x}_{c_i}$.
Note that the coefficients $\{\gamma^1_{c_i},..., \gamma^K_{c_i}\}$ are learnable parameters that are initialized as zero.

This procedure can be written as follows:

\begin{equation}
\hat{x}^j_{c_i} = \sigma (\hat{x}_{c_i} + \gamma^{j}_{c_i}g_{\theta}(z^j_{c_i}, w))
\end{equation}

Here, $\sigma$ is an activation function used to maintain the solution candidates within the data distribution and $w$ is a random noise sample drawn from a normal distribution. We set the data range to $[-1,1]$ and employ the $tanh$ function as $\sigma$. By introducing $\sigma$, there is no need to use the boundary loss $\mathrm{L}_{\mathrm{bnd}}$ in the NA method.

The aforementioned procedure can be contrasted with the \textit{mutation} process in the genetic algorithm\cite{holland1992adaptation}. However, although mutation directly perturbs the candidates, the NeuLag method employs latent variables and a network to perturb $\hat{x}_{c_i}$. This approach was adopted for three main reasons.

The first reason is the difficulty in adjusting the scale of the perturbations in the input space $X$. Directly perturbing the solution candidate $\hat{x}_{c_i}$ may worsen the solution quality if the perturbation scale is too large. Unlike GA, which improve solution candidates by repeatedly mutating them, the NeuLag method only branches out the solution candidates once, and the remainder of the optimization is performed using backpropagation. This makes it difficult to adjust the scale. Therefore, we delegate the scale adjustment to the neural network using scale factors. The learnable scale factor $\gamma$ is initially set to zero and optimally adjusted by backpropagation.
Second, we believe that adding noise prevents the solution candidates from converging to the same optimal solution. Candidates that branched from the same solution $\hat{x}_{c_i}$ are likely to converge to the same optimal solution. We prevent this by adding random noise to the network and encourage it to search further for the optimal area. 
Finally, optimization using a noisy network allows us to sample the solution. Our projector network $g$, which is similar to StyleGAN, adds noise to each intermediate layer for each propagation. With this system, once optimized, we can sample a large number of solutions simply by iteratively propagating the projector network $g$.

In this study, we constructed a projector network denoted as $g$, defined by the mapping 
$g: Z \rightarrow X$, where $Z$ is the latent space and $X$ is the input space of the surrogate simulator. This projector network is composed of several residual blocks, each incorporating layer normalization\cite{ba2016layer} and swish activation\cite{ramachandran2017searching}. Across all three AEM tasks, a consistent configuration of four residual blocks was employed. Notably, the total number of parameters for this network is 730K, which is considerably lower in comparison to the medium surrogate simulator used in all three tasks. For more details, please refer to Section \ref{sec:proj}.

\begin{figure}[t]
\centering
  \includegraphics[width=\textwidth]{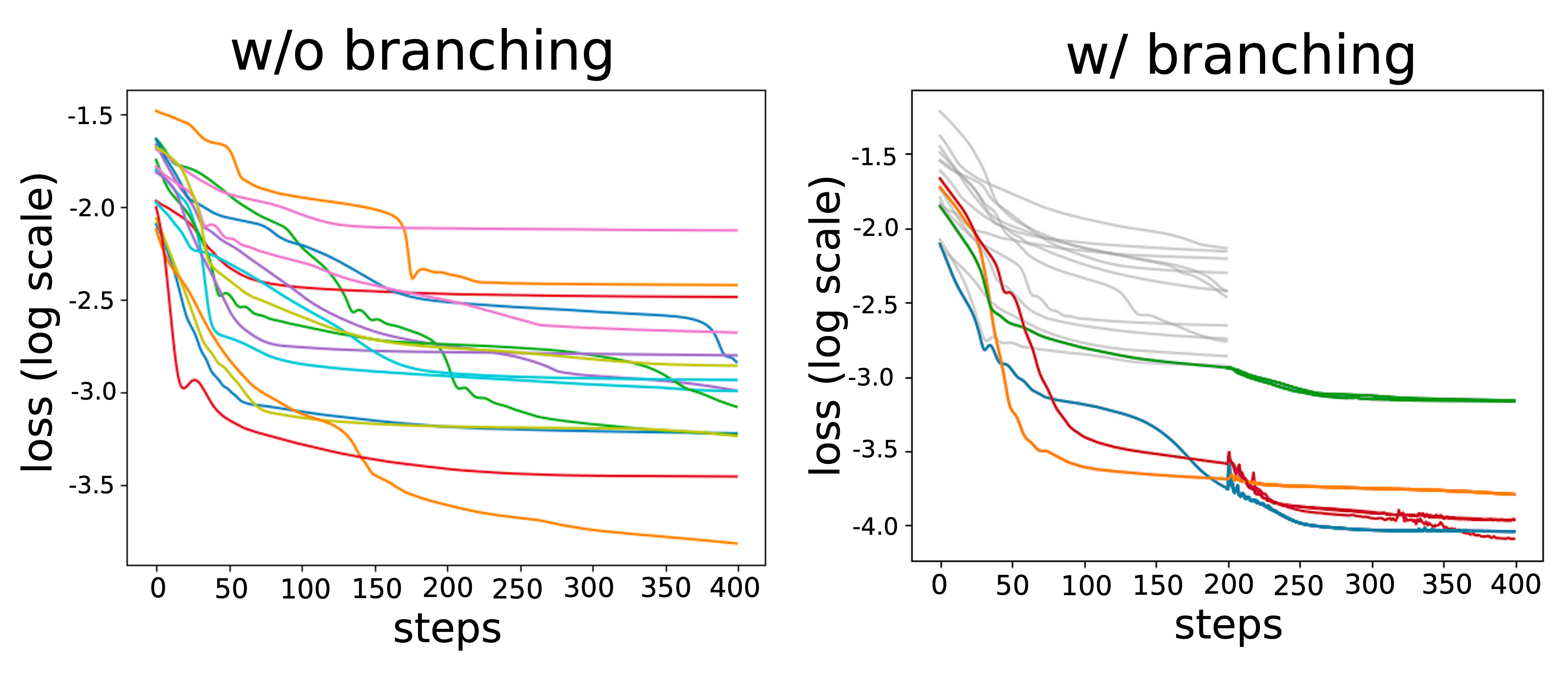}
  \caption{Branching out and selection for the promising solution candidates with batch size = 16, number of promising candidates $N=4$, and number of branches $K=4$. Computational resources can be focused on the most promising candidates by selecting and branching out promising solution candidates. (left) w/o branching out and selection. (right) w/ branching out and selection. At 200 steps, we reduce the number of candidates to 1/4 and branch them out. The same color indicates branches from the same solution candidate. Red candidates converge to different optima due to branching out. Gray lines indicate unpromising candidates}
  \label{fgr:branching_selection}
\end{figure}

In practice, a batch is composed of multiple solution candidates and optimization is performed in parallel for each candidate within the batch, similar to the NA method. For example, suppose that the batch size is 32, the number of promising candidates $N$ is 4, and the number of branches $K$ is 8. In this case, the NeuLag method uses randomly initialized latent variables $\{ z^{1}_{c_1},...,z^{8}_{c_1},...,z^{8}_{c_4} \}$ and coefficient $\{ \gamma^{1}_{c_1},...,\gamma^{8}_{c_1},...\gamma^{8}_{c_4} \}$. 
The number of latent variables and coefficients is the same as the batch size (Fig \ref{fgr:branching_selection}).

The loss in the NeuLag method, $\mathrm{L}_{\mathrm{NL}}$, is calculated similarly to that in the NA method, except we use the activation function $\sigma$ instead of $\mathrm{L}_{\mathrm{bnd}}$:

\begin{equation}
\hat{f_s}(\hat{x}^j_{c_i}) = \hat{f}_s(\sigma (\hat{x}_{c_i} + \gamma^{j}_{c_i}g_{\theta}(z^j_{c_i}))) = \hat{y}
\end{equation}

\begin{equation}
\mathrm{L}_{\mathrm{NL}}= \mathrm{MSE}(\hat{y}, y_{gt})
\end{equation}

Although we use an additional projector network, there may be concerns about its impact on the optimization speed compared to the NA method. However, when employing a highly accurate surrogate simulator, the projector size is small compared to the size of the surrogate simulator; therefore, the speed is not significantly affected. The optimization speed is discussed in Section \ref{experiments}.

\subsection{Introducing constraints}
\label{intro_const}

It is noteworthy that we can extend $\mathrm{L}_{\mathrm{bnd}}$ of the NA method, thereby introducing a soft constraint to both the NA and NeuLag methods. Additionally, we can introduce hard constraints, which have not been investigated using the NA method. In the subsequent sections, we provide a detailed explanation of the process for introducing soft and hard constraints to both the NA and NeuLag methods.

First, \textit{soft} constraints are considered.
Because the NeuLag method uses an activation function $\sigma$ to maintain the solution candidates in the data range, we do not need to introduce a \textit{soft} constraint for this purpose, unlike the NA method. Nevertheless, if necessary, the following constraints can be introduced.

\begin{equation}
\mathrm{L} = \mathrm{L}_{\mathrm{MSE}} + \lambda \mathrm{L}_{\mathrm{cons}}
\end{equation}

where $\mathrm{L}_{\mathrm{cons}}$ is a loss representing a constraint, and $\lambda$ is a constant indicating the strength of the constraint. We can obtain solutions that satisfy this constraint by minimizing the loss function. Drawing upon the structural resemblance to the \textit{Lagrangian multiplier}, we named our method Neural Lagrangian, abbreviated as, NeuLag.

In addition to the soft constraint, we can consider \textit{hard} constraints.
In hard constraints, some features of the solution candidates are forced to a specific value, for example, optimizing $x_0$, $x_2$, and $x_3$, while fixing $x_1$ to a specific value. In this case, the obtained solution always satisfies the constraint, unlike the soft constraint.

These methods for introducing soft or hard constraints are applicable to both the NA and NeuLag methods. In this study, we examined whether the NA and NeuLag methods can obtain reasonable solutions under soft or hard constraints.

\section{Experiments}
\label{experiments}

This section presents the experimental results of the NeuLag and NA methods. We describe the experimental setup and surrogate simulators in Section 4.1. In Section 4.2, we discuss our findings on the relationship between the accuracy of the surrogate simulator and the quality of the inverse-problem solutions. Next, we compare our results with those of other methods in Section 4.3, and present the optimization results under soft and hard constraints in Section 4.4. Finally, we examine the effectiveness of the NeuLag method in situations where the batch size is limited due to the use of a large surrogate simulator in Section 4.5.

\subsection{Experimental settings}

\subsubsection{Evaluation metrics}

We adopted the same evaluation metrics as those of Ren \text{et al.} for the inverse problem\cite{ren2022inverse}. The quality of the solutions obtained using DIMs was evaluated using the distance between the ground-truth simulation output $f(\hat{x_{c}})$ and target spectra $y_{gt}$. This evaluation metric is called the resimulation error \cite{kruse2021benchmarking} and is defined as 
Equation (\ref{resim_error}).

However, performing evaluations using simulator $f$ for all solution candidates is costly. The performance of DIMs strongly depends on the number of samples evaluated using simulator $f$. Therefore, Ren \textit{et al.} \cite{ren2020benchmarking} proposed an evaluation with $f$ only for the top-T solution candidates, in which the losses of the surrogate simulator $\hat{f}_s$ are small, and chose the best candidate. This evaluation metric is denoted as $r_T$. For example, with $r_{T=200}$, the evaluation is performed by selecting the top-200 solution candidates whose losses calculated with the surrogate simulator $\hat{f}_s$ are the first to the 200-th smallest among the solution candidates, and choosing the best candidate. The evaluation metric using dataset $D$ can be expressed as follows:

\begin{equation}
r_T = \mathbb{E}_{y_{gt} \sim D}\left[
 \min_{i \in [1,T]}\mathrm{L}_{\mathrm{resim}}(f(\hat{x_i}),y_{gt}) \right]
\end{equation}

We used 1 and 200 as the T values, according to Ren \textit{et al.}\cite{ren2022inverse}. The value $r_{T=1}$ is often adopted in the DIMs evaluation in AEM. Because the performance of DIMs improves as T increases, we used $r_{T=200}$ as the asymptotic evaluation for $T \rightarrow \infty$.

\subsubsection{The datasets}
\label{dataset}

In this paper, three AEM datasets were used: Stack \cite{chen2019smart}, Shell \cite{peurifoy2018nanophotonic}, and ADM \cite{deng2021neural}, as done by Ren \textit{et al.}\cite{ren2022inverse}. These datasets come from different subfields, and are thus expected to enable us to assess the overall performance of the NeuLag method on the AEM using these datasets (Table~\ref{tbl:dataset}).  

The task of the \textit{Stack} dataset is to optimize the geometry of a multilayer stack of alternating graphene and Si$_3$N$_4$ dielectric layers to produce the target absorption spectra under $s$-polarized light. The \textit{Shell} dataset considers the TiO$_2$–silica multilayer shell, a spherical nanoparticle with alternating layers of TiO$_2$ and silica. The task of this dataset is to optimize the geometry to produce the target scattering-cross-section spectra. We parameterize the thickness of TiO$_2$ and the silica shells of the nanosphere. The task of the \textit{ADM} dataset is optimizing the 14-dimensional geometry inputs of an all-dielectric metasurface supercell consisting of four SiC elliptical resonators to produce a target absorptivity spectrum. This dataset was selected because of its complexity and sharp peaks in its scattering response.

\begin{table}[t]
\centering
\small
  \caption{ Statistics about datasets. Dataset size in this table is for the \textit{base} and the \textit{medium} surrogate simulator}
  \label{tbl:dataset}
  \begin{tabular*}{0.6\textwidth}{@{\extracolsep{\fill}}cccc}
    \hline
     & Stack & Shell & ADM\\
    \hline
    description & multilayer stack & multilayer shell & metasurface \\
    $x$ dimension & 5 & 8 &  14\\
    $y$ dimension & 256 & 201 & 2000 \\
    dataset size & 50,000 & 50,000 & 10,000\\
    \hline
  \end{tabular*}
\end{table}

\subsubsection{Software and hardware}
The software applications used were PyTorch \cite{Paszke_PyTorch_An_Imperative_2019} and PyTorch Lightning \cite{Falcon_PyTorch_Lightning_2019}. We used an NVIDIA A100 GPU to accelerate the optimization. Our project was based on Ascender \cite{Fukuhara_Ascender_2022}.

\subsubsection{The Surrogate simulators}
\label{surrogate_simulator_explain}

We prepared three surrogate simulators, \textit{base}, \textit{medium}, and \textit{large}, to investigate the impact of surrogate simulator accuracy on solution quality.
The AEM tasks featured more output dimensions than input dimensions. These problem setups are similar to those of the image-generation tasks. 
Considering this similarity, we constructed network architectures with reference to the leading generative models, StyleGAN\cite{karras2019style} and BigGAN \cite{brocklarge}. 
These are generative methods, called generative adversarial networks (GANs) \cite{goodfellow2020generative}, which are unsupervised learning methods where two neural networks compete in a zero-sum game. 
Similar to StyleGAN, we employed a surrogate simulator network that accepts learnable parameters as inputs, and the surrogate simulator takes nonlinearly transformed features $x$ through AdaIN\cite{huang2017arbitrary}, which is a normalization layer that accepts two inputs. Finally, to avoid gradient vanishing, we used skip connections\cite{he2016deep} which are shortcut branches that skip some layers. For more details, please refer to Section~\ref{apdx:surrogate_simulators_explain}

According to the latest findings in deep learning, model accuracy improves as dataset and model sizes increase \cite{kaplan2020scaling, kolesnikov2020big, dosovitskiy2020image}. Therefore, we trained three surrogate simulators with different accuracies by progressively increasing the dataset and model size.

We trained the \textit{base} and \textit{medium} surrogate simulators by increasing only the model size while maintaining the dataset size constant, following the approach of Ren \textit{et al.} \cite{ren2022inverse} for comparison with other methods. We also trained the \textit{large} surrogate simulator using four times as much data as that used for \textit{base} and \textit{medium}, while simultaneously increasing the model size to improve the surrogate simulator’s accuracy. 

The accuracies of the surrogate simulators are listed in Table \ref{tbl:surrogate_sim_acc}. The "ref" in this table indicates the surrogate simulator accuracy of the NA method reported by Ren \textit{et al.}\cite{ren2022inverse} \footnote[1]{\url{https://github.com/BensonRen/AEM_DIM_Bench/tree/main/NA/models}}.

\begin{table}[t]
\centering
\small
  \caption{ Validation losses of surrogate simulators and train dataset sizes. }
  \label{tbl:surrogate_sim_acc}
  \begin{tabular*}{0.6\textwidth}{@{\extracolsep{\fill}}ccccc}
    \hline
     & ref & base (ours)& medium (ours)& large (ours)\\
    \hline
    Stack & 2.31e-6 & 3.67e-6 & 3.90e-7 & \textbf{7.11e-8}\\
    Shell & 5.61e-3 & 2.33e-3 & 4.92e-4 & \textbf{4.46e-5} \\
    ADM & 1.29e-3 & 3.65e-3 & 1.71e-3 & \textbf{3.63e-4} \\
    \hdashline
    dataset size & $\times1$ & $\times1$ & $\times1$ & $\times4$  \\
    \hline
  \end{tabular*}
\end{table}

\subsection{Model size vs resimulation error}

In this section, we investigate the impact of the surrogate simulator’s accuracy on the quality of the solutions to the inverse problems.
To do this, we solved the inverse problem using three surrogate simulators (\textit{base}, \textit{medium}, and \textit{large}) and measured the loss of the forward model $\mathrm{L}_{\mathrm{FW loss}}$ and the true simulator $\mathrm{L}_{\mathrm{resim}}$.
In this experiment, the batch size was set to 2048. The number of optimization steps was 300. Both the NA and NeuLag methods processed a batch size consisting of 2048 solution candidates during each optimization step. Note that both methods collectively processed a total of 2048 $\times$ 300 solution candidates throughout the optimization. In all experiments, the projector networks $g$ were the same and consisted of fully connected layers with layer normalization\cite{ba2016layer} and skip connections. The number of solution candidates optimized simultaneously was 2048 for the NA method and 8192 for the NeuLag method. The NA method could only optimize that batch size of candidate solutions simultaneously. However, as explained in Section 3, the NeuLag method focuses computational resources on promising solution candidates, allowing us to handle many solution candidates with the same computational resources and memory sizes.

\begin{figure}[t]
\centering
  \includegraphics[width=0.75\textwidth]{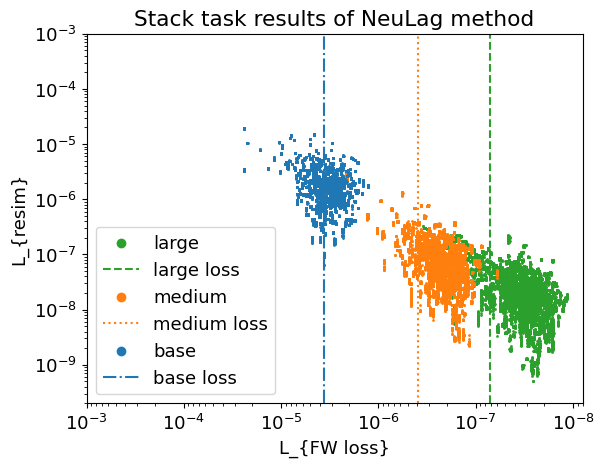}
  \caption{Relation between the loss of surrogate simulators ($\mathrm{L}_{\mathrm{FW loss}}$) and resimulation error ($\mathrm{L}_{\mathrm{resim}}$) for results of Stack task using NeuLag method. Three surrogate simulators with different accuracy, \textit{base}, \textit{medium}, and \textit{large}, were used in the experiments. Each point represents solutions, and dashed line represents the validation loss of surrogate simulators.}
  \label{fgr:acc_vs_inv_neulag}
\end{figure}

The qualitative results of the NeuLag method are shown in Fig.~\ref{fgr:acc_vs_inv_neulag}. The quantitative evaluation results are listed in Table \ref{tbl:surrogate_vs_inv_acc}. As displayed in Fig.~\ref{fgr:acc_vs_inv_neulag}, the upper bounds of the inverse-problem solution quality (lower bound on the resimulation error $\mathrm{L}_{\mathrm{resim}}$) are located around the validation loss of each surrogate simulator, similar to NA method (Fig.~\ref{fgr:acc_vs_inv}). Although the number of plotted data points is same in both Fig.~\ref{fgr:acc_vs_inv} and Fig.~\ref{fgr:acc_vs_inv_neulag}, Fig.~\ref{fgr:acc_vs_inv_neulag} appears to have fewer data points.
This is because the NeuLag method branches out a single candidate solution into multiple solutions; thus, the solutions converge at similar locations, and the data points overlap. Additionally, Table \ref{tbl:surrogate_vs_inv_acc} shows that the solution quality improved with the accuracy of the surrogate simulator. Therefore, the accuracy of the surrogate simulator significantly affects the accuracy of the inverse problem.

In Fig.~\ref{fgr:acc_vs_inv}, the resimulation error tends to be lower than the forward loss. This is because the forward loss of the candidate solutions was reduced to the limit of the surrogate model's accuracy, which increased the probability of falling into areas with small resimulation errors. The resimulation error was agnostic to the surrogate model. The surrogate simulator could improve the resimulation error of the solution candidates with forward losses larger than the validation error because they are in areas where the approximation of surrogate simulator is better. However, solution candidates with forward losses less than or equal to the validation error could no longer be evaluated using the surrogate model. Therefore, we considered these solution candidates to have a higher probability of having a resimulation error value lower than the forward loss.

\begin{table}[t]
\centering
\small
  \caption{ Results of $r_{T=1}$ evaluation for the Stack, Shell, and ADM tasks. The NA and NeuLag methods shared the same surrogate simulator. The score is averaged over the evaluation for 500 distinct test data samples. }
  \label{tbl:surrogate_vs_inv_acc}
  \begin{tabular*}{0.60\textwidth}{@{\extracolsep{\fill}}cccccc}
    \hline
     & & reference\cite{ren2022inverse} & base & medium & large\\
    \hline
    & val-loss & 2.31e-6 & 3.67e-6 & 3.90e-7 & \textbf{7.11e-8} \\
    
    \cdashline{2-6}
    Stack& NA & 1.22e-6 & 1.28e-6 & 1.03e-7 &2.86e-8\rule[0mm]{0mm}{3mm} \\
    & NeuLag & - &1.27e-6&1.28e-7&\textbf{1.27e-8}\\
     \hline
    & val-loss &5.61e-3&2.33e-3&4.92e-4& \textbf{4.46e-5} \\
    
    \cdashline{2-6}
    Shell& NA &3.60e-3&8.31e-4&1.54e-4&\textbf{1.23e-4}\rule[0mm]{0mm}{3mm} \\
    & NeuLag & - &8.49e-4&1.79e-4&1.26e-4\\
    \hline
    & val-loss &1.29e-3&3.65e-3&1.71e-3& \textbf{3.63e-4} \\
    
    \cdashline{2-6}
    ADM& NA &1.16e-3&4.35e-3&5.81e-4&2.56e-4\rule[0mm]{0mm}{3mm} \\
    & NeuLag & - & 4.32e-3&3.52e-4&\textbf{2.51e-4}\\
    \hline
  \end{tabular*}
\end{table}

\begin{table}[t]
\centering
\small
  \caption{Optimization speed comparison with the Stack task. Optimization steps are 300 for both with an NVIDIA A100 GPU.}
  \label{tbl:speed }
  \begin{tabular*}{0.35\textwidth}{@{\extracolsep{\fill}}cccc}
    \hline
     & base & medium & large\\
    \hline
    NA (s) & 17.56 & 31.03 &88.97\\
    NeuLag (s) & 20.79& 34.20 &91.86\\
    \hline
  \end{tabular*}
\end{table}

\begin{table*}
\small
  \caption{\ Comparison with other DIMs. Note that $\ast$ indicates a different training data size for surrogate simulator. Scores of other methods are quoted from the paper\cite{ren2022inverse}.} 
  \label{tbl:comparison}
  \begin{tabular*}{\textwidth}{@{\extracolsep{\fill}}lwc{1.0cm}wc{1.0cm}wc{1.0cm}wc{1.0cm}wc{0.8cm}wc{0.8cm}wc{0.8cm}wc{0.8cm}wc{0.8cm}wc{0.8cm}wc{0.8cm}wc{0.8cm}}
    \hline
    &\begin{tabular}{c}NeuLag\\large$\ast$\\(ours)\end{tabular}&\begin{tabular}{c}NA\\large$\ast$\\(ours)\end{tabular}&\begin{tabular}{c}NeuLag\\medium\\(ours)\end{tabular}&\begin{tabular}{c}NA\\medium\\(ours)\end{tabular}&NA&GA&NN&TD&INN&cINN&MDN&cVAE\\
    \hline
    Stack& & & & & & & & & & & & \\
    $r_{T=1}$&\textbf{2.83e-8}&2.86e-8&1.28e-7&1.03e-7&1.22e-6&1.39e-6&6.37e-7&4.37e-6&1.30e-3&9.38e-6&3.95e-5&1.86e-6\\
    $r_{T=200}$&\textbf{1.06e-8}&2.14e-8&1.05e-7&6.35e-8&5.87e-7&1.36e-6&-&-&1.29e-3&3.78e-7&3.03e-7&2.46e-7\\
    \hline
    Shell& & & & & & & & & & & & \\
    $r_{T=1}$&1.26e-4&\textbf{1.23e-4}&1.79e-4&1.54e-4&3.60e-3&3.91e-3&6.12e-3&7.13e-3&1.08e-2&9.03e-2&1.02e-1 &1.09e-2\\
    $r_{T=200}$&1.19e-4&1.14e-4&1.48e-4&\textbf{8.36e-5}&1.58e-3&3.02e-3&-&-&9.36e-3&1.03e-2&8.05e-3&1.91e-3\\
    \hline
    ADM& & & & & & & & & & & & \\
    $r_{T=1}$&\textbf{2.51e-4}&2.56e-4&3.52e-4&3.81e-4&1.16e-3&1.10e-3&1.72e-2&1.66e-3&9.08e-3&7.45e-3&7.34e-3&6.42e-3\\
    $r_{T=200}$&1.67e-4&\textbf{1.27e-4}&2.79e-4&2.20e-4&3.00e-4&7.73e-4&-&-&9.05e-3&1.73e-3&1.55e-3&1.39e-3\\
    \hline
  \end{tabular*}
\end{table*}

Speed comparisons are presented in Table \ref{tbl:speed }. In this experiment, we performed 300 optimization steps of the Stack task. We set the number of solution candidates to 2048 for the NA method and 8192 for the NeuLag method. Because of the additional projector network and larger pool of candidates, the NeuLag method took longer to optimize than the NA method. However, the difference was approximately 3.2\% for the \textit{large} simulator because the projector's influence was relatively small. Therefore, we consider that the impact on speed was negligible. A comparison between the NA and NeuLag methods using the same number of solution candidates is presented in Section \ref{limitation}.

\subsection{Resimulation error comparison with other methods}

Table~\ref{tbl:comparison} presents the results of the comparison with other DIMs. We set the batch size to 512 for ADM - \textit{large} owing to the GPU memory limit and to 2048 for the others. The number of simultaneously optimized solution candidates was 512 for the NA method and 2048 for the NeuLag method in ADM - \textit{large}. For the other cases, we set it to 2048 for the NA method and 8192 for the NeuLag method. The number of optimization steps was 300 for Stack and ADM. For the Shell task, we set 1200 steps because it did not converge in 300 steps. We used the Adam optimizer\cite{kingma2014adam} with 0.01 learning rate for all tasks and for both the NA and NeuLag methods. We set the number of final promising candidates to 64, and the number of branches for each promising solution candidate to eight for all tasks.

In the Stack task, the value of the resimulation error was at most 1/5 compared to other methods for the same conditions of the surrogate simulator dataset. In the experiments with more training datasets for the surrogate simulator, the value of the resimulation was at most 1/50 that of the previous methods (NA-cVAE) in Table~\ref{tbl:comparison}. In Shell and ADM, the value of the resimulation was approximately 1/5 to 1/30 of that of the previous methods in the experiment using the \textit{large} surrogate simulator.

\subsection{Optimization under constraint}

In real-world problems, inverse problems often need to be solved under certain constraints. In some cases, it is necessary to solve inverse problems while satisfying specific constraints. For example, there is growing interest in Na-ion batteries as alternatives to Li-ion batteries (LiB). Li is a lightweight material with a large standard redox potential,  making it suitable for use in batteries with high energy densities and voltages. However, Li has cost and supply issues; therefore, researchers are exploring the use of Na as an alternative\cite{slater2013sodium}. This problem setup essentially involves solving an inverse problem of maximizing the energy density or voltage under the constraint of using Na instead of Li. Thus, verifying behavior under constraints is essential for inverse-problem methods in the real world. Here, we verify that the proposed method works well under certain constraints.
Although the constraints tested here are hypothetical and not directly connected to real problems, they are expected to be useful for examining the behavior of the solutions of DIMs under a constraint.

To verify the behavior of our method under a constraint, we experimented with two cases: \textit{soft} and \textit{hard} constraints. In the \textit{soft} constraint case, we imposed the constraint that the sum of certain features takes a constant value, that is, $|c_1 x_1 +c_2 x_2 + c_3 x_3 -a|=0$, where $x_1, x_2, x_3$ are features, and $a, c_1, c_2$  and $c_3$ are constants. In this case, the problem is reduced to the minimization of the following loss function:

\begin{equation}
\mathrm{L} = \mathrm{L}_{\mathrm{MSE}} + \lambda |c_1 x_1 +c_2 x_2 + c_3 x_3 - a|
\end{equation}

The NA method introduced a soft constraint in the form of a boundary loss. However, the constraint we used above is a more generalized form of boundary loss and has a wider range of applications.

Specifically, we randomly selected a maximum of three features. Subsequently, the three constants $c_1, c_2$ and $c_3$ are randomly selected from $0, \pm1, \pm0.5$ and $a$ were set to zero. We also set the strength of the constraint $\lambda$ to $1$ for all experiments. We prepared nine different constraints in this manner and solved the inverse problem by imposing them on 50 $y_{gt}$ spectra for each of the three AEM tasks. Finally, $r_T$ scores were computed for 450 experiments in each of the three AEM tasks to examine the behavior under a soft constraint.

In the second case, we considered a hard constraint that forces one feature to assume a specific value. Specifically, one feature was chosen from $x_0,...$ and forced its value to be $-1$, the minimum value of the feature; for example, $x_0$, $x_2$, and $x_3$ are subject to optimization, whereas $x_1$ is always set to $-1$. Python-like pseudocode is presented in Algorithm \ref{alg:hard}.

\begin{algorithm}[t]
\caption{hard constraint}
\label{alg:hard}
\begin{algorithmic}[1]
\renewcommand{\algorithmicrequire}{\textbf{Input:}}
\renewcommand{\algorithmicensure}{\textbf{Output:}}
\REQUIRE  input: $x$, the number of features of $x$: $n$, the index of constraint feature: $f_i$, a constant value of hard constraint: $c$
\ENSURE hard constraint input: $x_{hard}$

\STATE $\mathrm{cons} = \mathrm{zero \_ tensor(shape}=(n,))$
\STATE $\mathrm{cons}[f_i] = c$
\STATE $\mathrm{used\_feture} = \mathrm{zero \_ tensor(shape}=(n,))$
\STATE $\mathrm{used\_feture}[f_i] = 1.0$
\STATE $\mathrm{used\_feture} = 1.0 - \mathrm{used\_feture}$
\STATE $x_{hard} = x * \mathrm{used\_feture} + \mathrm{cons}$
\end{algorithmic}
\end{algorithm}

We experimented with 50 spectra randomly sampled for each task. For each spectrum, we repeated the following experiment for all input features: selecting a feature to be fixed at $-1$ and solving the inverse problem. We then calculated the mean value of the resimulation error for all the experiments for each task. This setup can be applied, for example, to the development of Li-free batteries. Consider a problem setup in which each feature represents the ratio of an element that constitutes the battery; where the lowest value indicates that the element is not used at all. Therefore, solving the inverse problem of finding a battery material that maximizes energy density and voltage while fixing the Li feature to the lowest value (i.e., containing no Li) corresponds to the hard constraint problem.

 Table \ref{tbl:soft_onstraint_satisfuction} shows the satisfaction values of the soft constraints, which are calculated as the average of the absolute values of the constraint formula for the top-200 samples of $L_\mathrm{FW loss}$ (samples from the 1st to the 200th lowest $L_\mathrm{FW loss}$). A satisfaction value of zero indicates that the constraint was rigorously satisfied. It is worth noting that the constants $c_1, c_2$, and $c_3$ are in the range of $\pm 1$, the equation used at most three features, and the range of the data is [$-1,1$]. Therefore, the range of satisfaction values is [$0,3$]. Both the NA and NeuLag methods satisfied the constraints with a precision of approximately 1e-4 over the entire range.

The NeuLag method outperformed the NA method in terms of the satisfaction values of the Shell and ADM tasks. This is likely because the NeuLag method branches out only the solution candidate with the lowest loss, resulting in better average loss values for the top 200 samples. However, the NA method performed better on Stack task which is easier than the Shell and ADM tasks, as many solutions can reach a global optimum. Therefore, we speculate that the NA method may have been able to reach the global optimum for many solutions, whereas the NeuLag method was adversely affected by the noise added to the projector network.

\begin{table}[t]
\centering
\small
  \caption{Satisfaction values of the soft constraints, which are defined as average of the absolute values of constraint formula for top-200 samples that have the lowest $L_\mathrm{FW loss}$.}
  \label{tbl:soft_onstraint_satisfuction}
  \begin{tabular*}{0.3\textwidth}{@{\extracolsep{\fill}}ccc}
    \hline
    & NeuLag &  NA\\
    \hline
      Stack & 0.00329 &0.00192 \\
      Shell &0.00113&0.00208\\
      ADM &0.00081& 0.00329\\
    \hline
  \end{tabular*}
\end{table}

\begin{table*}
\small
  \caption{Results of optimization under constraint. Scores of other methods(NA, TD, INN, MDN) are quoted from the paper\cite{ren2022inverse}.}
  \label{tbl:constraint}
  \begin{tabular*}{\textwidth}{@{\extracolsep{\fill}}wc{0.8cm}wc{0.8cm}wc{0.8cm}wc{0.8cm}wc{0.8cm}wc{0.8cm}wc{0.8cm}wc{0.8cm}wc{0.8cm}wc{0.8cm}wc{0.8cm}wc{0.8cm}}
    \hline
    \multicolumn{5}{c}{w/ constraint} & \multicolumn{1}{c}{}&\multicolumn{6}{c}{w/o constraint}\\
    \cline{2-5} \cline{7-12}
     & & & & & & & & & & & \\
    & \begin{tabular}{c}NeuLag\\soft\end{tabular}&  \begin{tabular}{c}NA\\soft\end{tabular} &\begin{tabular}{c}NeuLag\\hard\end{tabular}& \begin{tabular}{c}NA\\hard\end{tabular}& & \begin{tabular}{c}NeuLag\\(ours) \end{tabular} & \begin{tabular}{c}NA\\(ours) \end{tabular}&NA&TD&INN&MDN\\
    \hline
    Stack & & & & & & & & & & & \\
    $r_{T=1}$&3.87e-5&\textbf{3.49e-5}&7.86e-5&7.86e-5& &1.27e-8&2.86e-8& 1.22e-6&4.37e-6&1.30e-3&3.95e-5\\
    $r_{T=200}$& 3.83e-5& \textbf{3.48e-5}& 7.84e-5  &7.84e-5 & &1.06e-8  &2.14e-8 &5.87e-7 &- & 1.29e-3 &3.03e-7  \\
    \hline
    Shell & & & & & & & & & & & \\
    $r_{T=1}$&2.74e-3&\textbf{2.17e-3}&1.58e-2&\textbf{1.39e-2}& &1.26e-4& 1.23e-4&3.60e-3&7.13e-3&1.08e-2&1.02e-1\\
    $r_{T=200}$ &2.74e-3&\textbf{2.12e-3}& 1.57e-2 &\textbf{1.32e-2} & &1.19e-4 &1.14e-4  &1.58e-3 &- &9.36e-3 & 8.05e-3\\
    \hline
    ADM & & & & & & & & & & & \\
    $r_{T=1}$&6.01e-4&\textbf{5.99e-4}&1.32e-3 &\textbf{1.19e-3} & &2.51e-4 & 2.56e-4&1.16e-3&1.66e-3& 9.08e-3&7.34e-3\\
    $r_{T=200}$ &4.91e-4&\textbf{4.38e-4}& 1.19e-3&\textbf{1.01e-3}& &1.67e-4 &1.27e-4 & 3.00e-4& -&9.05e-3 &1.55e-3 \\
    \hline
  \end{tabular*}
\end{table*}

The optimization results under soft and hard constraints are listed in Table \ref{tbl:constraint}. It is important to note that the existence of solution for this optimization problem under these constraints remains unclear. This suggests that the present optimization is a more difficult problem than an optimization without constraints. Therefore, significantly larger resimulation errors $r_{T=1}$ and $r_{T=200}$ under constraints than under no constraints would be acceptable. This may be justified by the fact that the resimulation errors were lower than those of some previous studies with no-constraint cases.

In most cases, the NA method performed slightly better than the NeuLag method. The NeuLag method optimizes solutions within certain ranges owing to the addition of noise, whereas the NA method optimizes at a point. Therefore, it is believed that the NA method slightly outperformed the NeuLag method in these constrained tasks, where the optimal areas were narrower than those in the non-constrained tasks.

Because this experiment does not directly reflect a real-world problem, no further discussions on the accuracy are possible. However, because it exhibits smaller resimulation errors than some previous methods, such as TD, INN, and MDN, as shown in Table \ref{tbl:constraint}, we expect that the NA and NeuLag methods have a high potential for solving optimization problems with soft or hard constraints.

\subsection{Advantages of NeuLag method with highly accurate surrogate simulators}

The NA method requires a new optimization round starting from the beginning to sample more than a batch of solution candidates, whereas the NeuLag method enables rapid sampling once projector optimization is performed.
This feature of the NeuLag method is advantageous when a highly accurate surrogate simulator is used because, in such a case, the model tends to be large; consequently, the batch size tends to be small.
This advantage is significant because several studies have claimed that deep-learning models become more accurate as the model size increases \cite{kaplan2020scaling, kolesnikov2020big, dosovitskiy2020image}. 

Here, we demonstrate the capability of the NeuLag method to alleviate the trade-off between solution quality and computational cost for inverse problems when using a highly accurate large surrogate simulator. Specifically, we use a \textit{large} simulator, a highly accurate large surrogate simulator, and a batch size of 16 to optimize the inverse problem.

This experiment used $r_{T=4}$ as the evaluation metric, assuming a situation in which the ground-truth simulator is computationally intensive, making it difficult to use a large T value to obtain better solutions. 
The solution candidates were obtained for the respective batches and then the obtained solution candidates were combined and evaluated using $r_{T=4}$. There were 16 solution candidates for the NA method (the same as the batch size) and 128 for the NeuLag method. The number of optimization steps was set to 300 for the NA method and 450 for the NeuLag method for the Stack and ADM tasks, while for the Shell task, the optimization step was set to 1200 for the NA method and 1800 for the NeuLag method. To evaluate more solution candidates simultaneously, the NeuLag method required many steps. Note that the cost of computing all solution candidates was approximately $1/43$ that of the NA method for the NeuLag method.

For the NA method, the optimization of the batch size, that is, 16 solution candidates, was repeated until the number of solutions reached 128. In the NeuLag method, 128 optimized solutions were sampled after optimization.

\begin{figure}[t]
\centering
  \includegraphics[width=\textwidth]{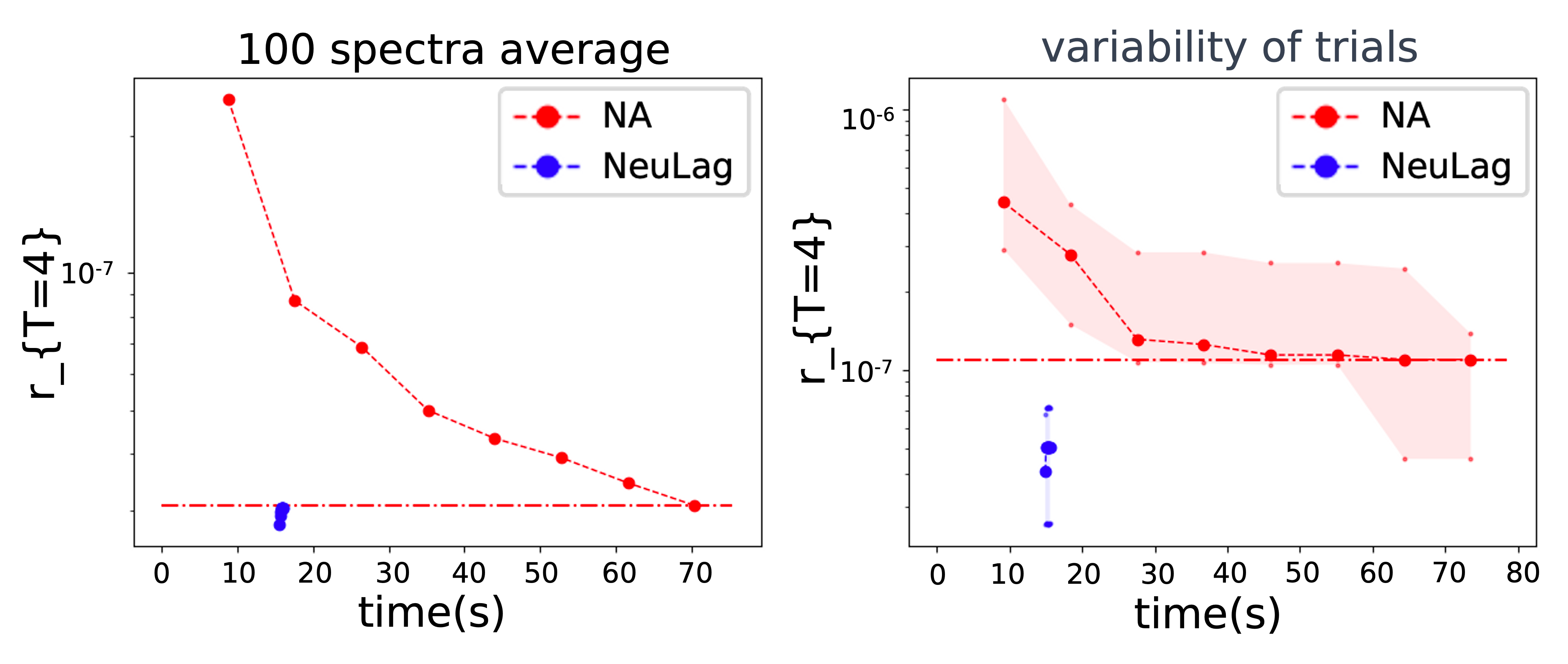}
  \caption{Relationship between sampling time and resimulation error with the Stack task. Dashed line indicates the best error of NA method. (left) Averaged results of 100 spectra. (right) Results of 10 trials with a spectrum. The dots indicate median values and colored areas indicate interquartile range (IQR) of 10 trials.}
  \label{fgr:sampling_time}
\end{figure}

The relationship between the time taken for sampling and the solution quality $r_{T=4}$ is shown in Fig \ref{fgr:sampling_time}. The NA method is fast in obtaining the first solution candidates, whereas the convergence speed is relatively low. However, the NeuLag method is slow in obtaining the first solutions, but its convergence speed is relatively high because of its fast sampling. In addition, the NeuLag method consistently yielded better solutions in all ten trials, whereas the NA method was unstable.

Table \ref{tbl:sampling_speed} lists the accuracy and sampling time for $r_{T=4}$ for the 128 samples. The NeuLag method could sample more accurate solutions than the NA method and 3.5 to 5.0 times faster. The resimulation error for the ADM task in NeuLag was worse than that of the NA method. However, the solution quality predicted by the surrogate simulator ($\min( L_{\mathrm{FW loss}})$) was almost identical. In the ADM task, both methods successfully optimized the solution quality to the upper bound of the accuracy of the large surrogate simulator (validation loss:3.63e-4), and the ground-truth simulator was agnostic of the surrogate simulator. Therefore, we consider that the difference in $r_{T =4}$ is coincidental and that the performances of the NA and NeuLag methods in the ADM task are equivalent.  

\begin{table}[t]
\centering
\small
  \caption{According to actual computational measurements, the total sampling time required for the optimization and sampling of 128 solution candidates is presented, along with the resimulation error $r_{T=4}$ and the minimum surrogate simulator loss among the 128 solutions that were optimized.}
  \label{tbl:sampling_speed}
  \begin{tabular*}{0.48\textwidth}{@{\extracolsep{\fill}}cccc}
    \hline
     & $r_{T=4}$ & $\min( L_{\mathrm{FW loss}})$ & sampling time (s) \\
    \hline
     Stack - NeuLag & 3.05e-8 & 2.96e-8 & 16.01 \\
     Stack - NA & 3.08e-8 & 2.54e-8 &70.39 \\
    \hline
     Shell - NeuLag & 1.95e-4 &1.95e-4 & 54.23 \\
     Shell - NA & 3.94e-4 & 3.94e-4&  276.16 \\
    \hline
     ADM - NeuLag & 2.44e-4 & 1.57e-4 & 14.94\\
     ADM - NA & 1.71e-4 & 1.50e-4 &  74.32 \\
    \hline
  \end{tabular*}
\end{table}

\section{Limitations}
\label{limitation}
Although we speculated that the accuracy of the surrogate simulator may ultimately limit the quality of the inverse-problem solutions obtained using the NA and NeuLag methods. However, even when using highly accurate surrogate simulator, training alone may not be sufficient to improve the quality of the inverse-problem solutions. This is evident from the resimulation errors of Shell-\textit{large} in Table~\ref{tbl:comparison}, unlike Stack, which do not reach the level of accuracy achieved by the surrogate simulator validation loss. Therefore, it is possible that the NA and NeuLag methods may not be able to achieve the quality limit of the inverse-problem solutions, which is determined by the accuracy of the surrogate simulator.

The NeuLag method is effective in selecting promising solution candidates when computational resources and memory sizes are limited, as shown in Table~\ref{tbl:sampling_speed}. However, with sufficient computational resources and memory sizes, it might be as effective as the NA method. Table \ref{tbl:same_candidates} presents the results of a 300-step optimization using the same number of candidate solutions as the NA method. In the NeuLag method with 2048 solution candidates, the number of candidate solutions was reduced to 32 in 200 steps, and branching started.
In this case, the results of the NeuLag method were slightly worse than those of the NA method.
This may be due to an inappropriate solution selection schedule in which the most promising solution candidates were truncated during the optimization process.

\begin{table}[t]
\centering
\small
  \caption{ Number of solution candidates vs resimulation error $r_{T=1}$ with Stack task using \textit{large} surrogate simulator. $n_c$ means the number of solution candidates. }
  \label{tbl:same_candidates}
  \begin{tabular*}{0.48\textwidth}{@{\extracolsep{\fill}}cccc}
    \hline
    method & NeuLag & NeuLag&NA\\
    $n_c$ & 8192$\rightarrow$ 32&2048$\rightarrow$ 32& 2048 \\
    \hline
     $r_{T=1}$ & 2.83e-8 & 3.61e-8 & 2.86e-8\\
    \hline
  \end{tabular*}
\end{table}

\section{Conclusion}

In this study, we investigated the effect of surrogate simulators accuracy on the solution quality for inverse problems. It was found that a higher accuracy in the surrogate simulator led to better solution quality. To address that challenge of optimizing a sufficient number of solution candidates using a lightweight projector network, we propose the NeuLag method. This method effectively searches for optimal areas by branching out the promising solution candidates. Furthermore, it can determine optimal solutions even when using a large and highly accurate surrogate simulator with a limited batch size. Our experiments on optimization under a constraint with NA and NeuLag also demonstrated their potential for use in solving problems with soft or hard constraints. We hope that this work will pave the way for real-world applications in areas such as molecules, materials \cite{hermann2020deep, pfau2020ab, behler2015constructing, schutt2017schnet}, and fluid problems \cite{raissi2019physics, mao2020physics} where complex surrogate simulators are used. We have released our codes to encourage further exploration of this field.

\section*{Acknowledgements}
We would like to express our sincere gratitude to Mr. Koji Fujii, Dr. Hideki Nakayama and Dr. Hirokatsu Kataoka for stimulating discussion. We also would like to thank Editage (www.editage.com) for English language editing.

\bibliography{references}{}
\bibliographystyle{hplain}

\newpage

\appendix
\renewcommand{\thesection}{A.\arabic{section}}
\renewcommand{\thesubsection}{\thesection.\arabic{subsection}}

\section{Surrogate simulators}
\label{apdx:surrogate_simulators_explain}

\begin{figure*}[h]
\centering
  \includegraphics[width=\textwidth]{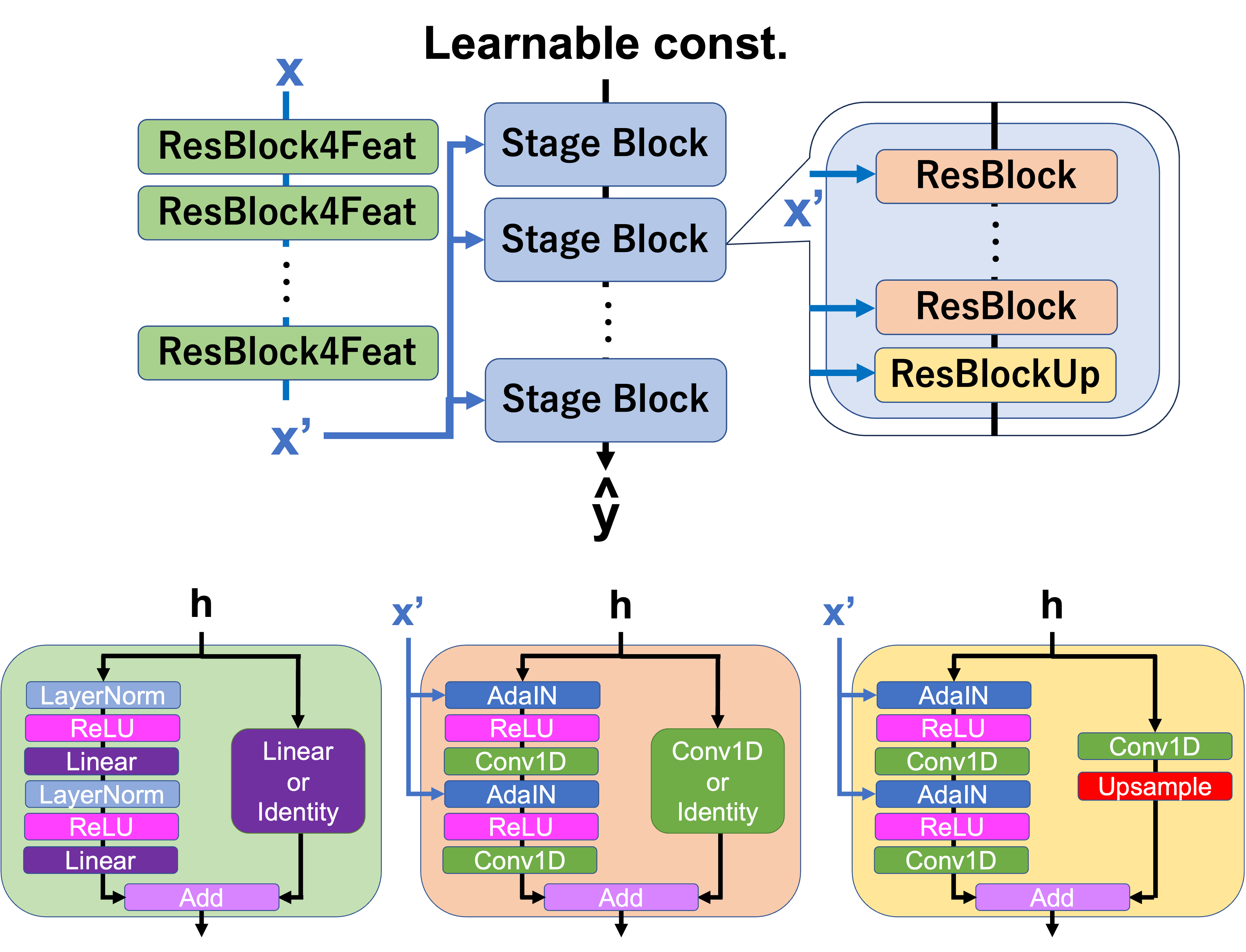}
  \caption{Architecture of the surrogate simulator. (top) Overview of the surrogate simulator. The feature vector $x$ is processed using ResBlock4Feat to perform a non-linear transformation, yielding $x'$. Subsequently, $x'$ is input into a Stage Block comprising several residual blocks\cite{he2016deep}. The mainstream of this block, similar to StyleGAN\cite{karras2019style} , receives a trainable tensor as input. (bottom left) ResBlock4Feat: Composed of linear transformations, layer normalizations\cite{ba2016layer}, and ReLU activations. If input and output dimensions are consistent, identity transformation is applied; otherwise, a linear transformation is used in the shortcut branch. (bottom center) ResBlock: Consists of AdaIN\cite{huang2017arbitrary}, ReLU, and Conv1D (1D convolution). Identity transformation is applied when input and output dimensions are the same; otherwise, Conv1D is used in the shortcut branch. (bottom right) ResBlockUp: Essentially identical to ResBlock but incorporates upsampling (Upsample) in the shortcut branch. For the final Stage Block's ResBlockUp, the upsampling is adjusted to match the output dimensions; for all other instances, the dimensions are doubled.}
  \label{fgr:surrogate-arch}
\end{figure*}

Fig~\ref{fgr:surrogate-arch} presents the detailed architectural diagrams of the surrogate simulators employed in this study, highlighting their structural components, layers, and specific data flow. Additionally, Table~\ref{tbl:surrogate_simulator_params} enumerates the hyperparameters utilized in the simulators, providing insights into their scalability. We utilized the Adam\cite{kingma2014adam} to train all surrogate simulators.

\begin{table*}[b]
  \caption{Hyperparameters used for training the base, medium, and large simulators across the Stack, Shell, and ADM tasks. "Width of mainstream blocks": The length of the lists indicates the number of Stage Blocks. The elements of this list denote the widths of each Stage Block as presented in Fig~\ref{fgr:surrogate-arch}. "No. of ResBlocks": Represents the count of ResBlocks in each Stage Block. The length of the lists indicates the number of ResBlocks. In these Stage Blocks, ResBlockUp follows these ResBlocks."No. of ResBlocks4Feat": Refers to the number of ResBlocks4Feat in the branch that processes the feature vector, as depicted in Fig~\ref{fgr:surrogate-arch} (top). The length of the lists indicates the number of ResBlocks4Feats and the empty lists indicate that identity transformation is employed instead of using ResBlocks4Feat.} 
  \label{tbl:surrogate_simulator_params}
  \begin{tabular*}{\textwidth}{@{\extracolsep{\fill}}c|ccccc}
    surrogate simulator & \begin{tabular}{c}width of\\main stream\\ blocks\end{tabular} & \begin{tabular}{c}No. of \\ ResBlocks\end{tabular}& \begin{tabular}{c}No. of \\ ResBlocks4Feat\end{tabular} & \begin{tabular}{c}learning \\ rate\end{tabular} & batch size\\
    \hline
    Stack\ - \ \it{base}&[256, 128, 64]&[0, 0, 0]&[~] & 0.01 & 1024\\
    Stack\ - \ \it{medium}&[512, 256, 128, 64]&[0, 0, 1, 1]&[64, 64]& 0.01 & 64\\
    Stack\ - \ \it{large}&[1024, 512, 256, 128]&[1, 1, 1, 2]&[128, 128, 128]& 0.001 & 32\\
    \hdashline
    Shell\ - \ \it{base}&[256, 128, 64]&[0, 0, 0]&[~]& 0.001 & 1024\\
    Shell\ - \ \it{medium}&[512, 256, 128, 64]&[0, 0, 1, 1]&[64, 64]& 0.001 & 64\\
    Shell\ - \ \it{large}&[1024, 512, 256, 128, 64]&[0, 0, 2, 2, 2]&[128, 128, 128]& 0.01 & 32\\    
    \hdashline
    ADM\ - \ \it{base}&[256, 128, 64]&[0, 0, 0]&[~]& 0.01 & 1024\\
    ADM\ - \ \it{medium}&[512, 256, 128, 64]&[0, 0, 1, 1]&[128, 128]& 0.01 & 64\\
    ADM\ - \ \it{large}&[1024, 512, 256, 128, 64]&[0, 0, 2, 2, 2]&[128, 128, 128]& 0.01 & 32\\    
    \hline
  \end{tabular*}
\end{table*}

\section{Projector network}
\label{sec:proj}
\begin{figure*}[t]
\centering
  \includegraphics[width=0.95\textwidth]{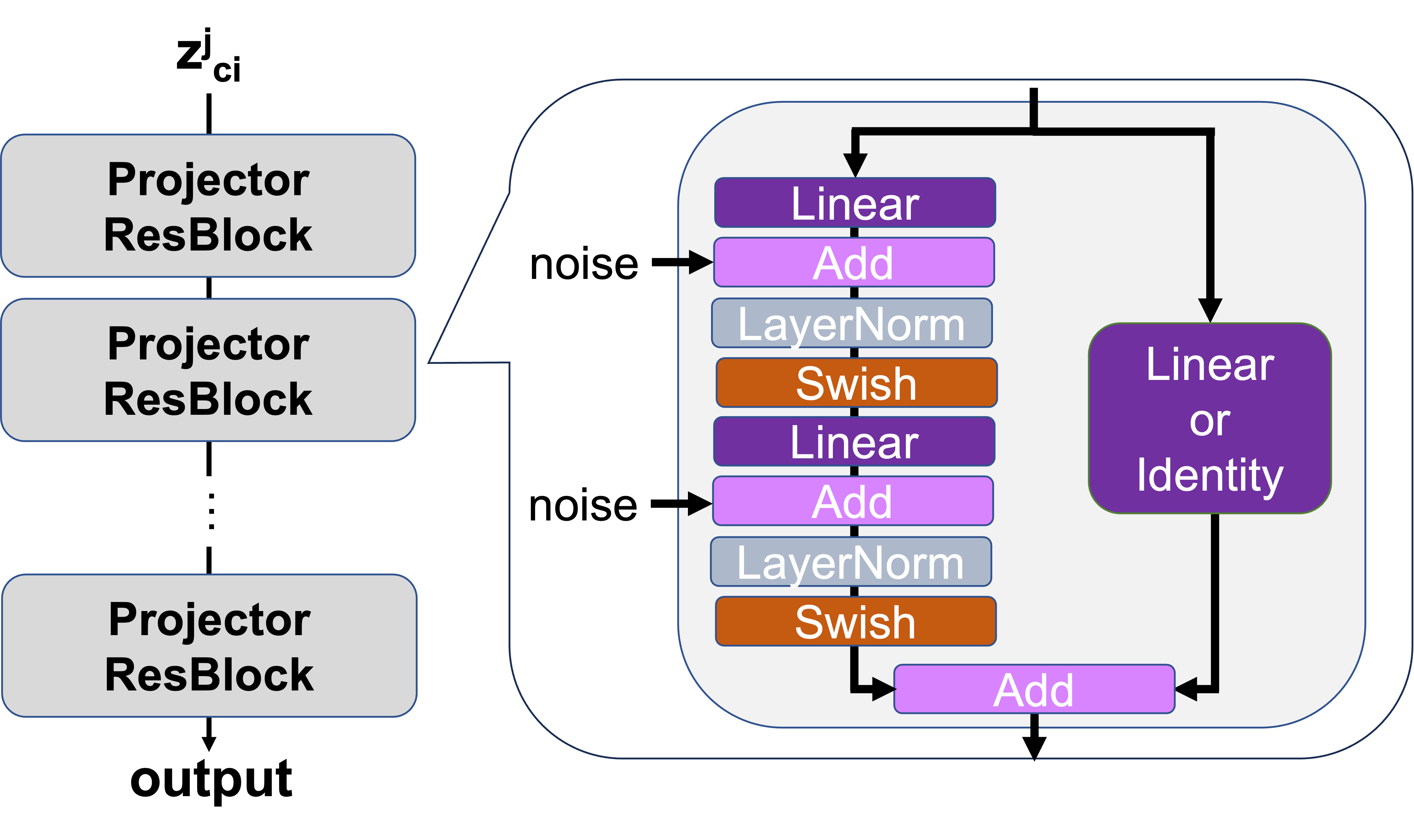}
  \caption{Architecture of the projector network. The network comprises several residual blocks. Each of these blocks incorporates linear transformation, layer normalization, and swish activation\cite{ramachandran2017searching}. On the right side of each residual block, the shortcut branch performs an identity transformation if the input and output dimensions are consistent; otherwise, it employs a linear transformation. Meanwhile, on the left side of the residual block, noises sampled from a Gaussian distribution are introduced after the linear transformation.}
  \label{fgr:proj}
\end{figure*}

Fig.~\ref{fgr:proj} illustrates the architecture of our projector network. Although the projector is built using multiple residual blocks, similar to the surrogate model, it is considerably more compact than the large surrogate simulators employed for each AEM task.

\end{document}